# Coupling silica waveguides to photonic crystal waveguides through multilayered Luneburg lens

S. Hadi Badri[a,*], M. M. Gilarlue[a]

[a] Department of Electrical Engineering, Sarab Branch, Islamic Azad University, Sarab, Iran

* sh.badri@iaut.ac.ir

**Abstract**

We present a detailed analysis of a coupler based on the Luneburg lens to couple a silica waveguide to a photonic crystal waveguide. The dependence of coupling efficiency on the lens's truncation, cut position of the photonic crystal structure, coupler tip width, and misalignment are investigated with two-dimensional finite element method. We implement the lens with a concentric ring-based multilayer structure. We also present a method to replace layers with very narrow widths by layers of predetermined minimum widths in the structure of the lens. The coupling loss of the designed 2.7 µm-long coupler, connecting a 2.79 µm-wide silica waveguide to a photonic crystal structure with a rod-type square lattice, is lower than 0.49 dB in the C-band. The average coupling loss in the entire S, C, L, and U bands of optical communications is 0.70 dB.

**Key words**

Luneburg lens; Photonic crystal waveguide; Silica waveguide; Optical coupler; All-dielectric Metamaterials

## 1. Introduction

Variety of key components have been designed by photonic crystals (PhCs) such as beam collimators [1, 2], lasers [3], slow-light waveguides [4], modulators [5, 6], topological insulators [7], and sensors [8]. In order to utilize the numerous devices based on PhCs and other planar lightwave circuits (PLC), efficient coupling of light into PhC devices is required. Coupling a silica waveguide to a PhC waveguide suffers from losses due to different guiding mechanisms, waveguide widths, as well as group index and modal mismatches in the waveguides. Propagation in PhC waveguides is characterized by Bloch modes while silica waveguides rely on total internal reflection [9, 10]. Variety of coupling mechanisms have been proposed to efficiently interface wide silica waveguides to narrow PhC waveguides. PhC step tapered coupler has been proposed to couple a 1.6 µm-wide ridge waveguide to a triangular PhC structure with a coupling length of 3 µm [11]. In this method, the minimum coupling loss is about 0.58 dB, however, the ripple in the transmission spectrum is considerably high. A 2.4 µm-long taper has been proposed to couple a 2.5 µm-wide silica waveguide into a rod-type square PhC structure with a minimum coupling loss of 0.46 dB [12]. A 2.3 µm-long coupler based on setting a single defect within a PhC taper has been reported where a 1.55 dB coupling loss has been achieved for coupling a 3 µm-wide ridge waveguide into a triangular PhC structure [13]. Defect-based PhC tapers have also been designed to couple silica waveguides into triangular rod-type PhC structures [14, 15]. Nonuniform PhC tapers with linear, convex, and concave curvatures have been studied [16]. The performance of linear dielectric taper, PhC taper, and graded PhC coupler, and parabolic mirror coupler have also been compared theoretically and experimentally [17].

Gradient index (GRIN) lenses such as Maxwell's fish-eye [18, 19], Luneburg [20, 21], and Eaton [22, 23] lenses have been used to design novel devices. In this paper, the focusing property of the Luneburg lens is utilized to couple a 2.79 µm-wide silica waveguide into a square PhC structure. The coupling efficiency is optimized by tuning the cut position of the photonic crystal structure and coupler tip width. The length of the designed coupler is 2.7 µm while the counterpart linear taper is considerably longer [17]. Numerical simulations indicate that the coupling loss is lower than 0.49 dB in the C-band while in the 1460-1675 nm bandwidth it is lower than 1.28 dB. The Luneburg lens is implemented by a concentric cylindrical multilayer structure where the feasibility of the design is increased by replacing layers with narrow widths by multiple layers with a predetermined minimum width. To the best of our knowledge, we present a dielectric waveguide to PhC waveguide coupler based on the truncated Luneburg lens for the first time.

## 2. Luneburg lens as coupler

The refractive index profile of the generalized Luneburg lens is described by [24]

$$n_{lens}(r) = n_{edge}\sqrt{1+f^2 - (r/R_{lens})^2}/f \quad , \quad (0 \leq r \leq R_{lens}) \tag{1}$$

where $n_{edge}$ is the refractive index of the lens at its edge, $r$ is the radial distance from the center, $R_{lens}$ is the radius of the lens, and $f$ determines the position of the focal point. The Luneburg lens focuses the parallel rays incident on its edge to a focal point determined by $f$. For $f=1$, the focal point lies on the edge of the lens while for $f<1$ or $f>1$ the focal point of the lens is located inside or outside of the lens, respectively. In our calculations we use $f=1$. As shown in Fig. 1, the ray-tracing calculations indicate that the Luneburg lens can couple a wider waveguide to a narrower waveguide. In this figure, the refractive index of the waveguides is 1.45. In order to minimize the reflections from the interface of the lens and the waveguides, the $n_{edge}$ should be the same as the refractive index of the waveguides, i.e., $n_{edge} = 1.45$.

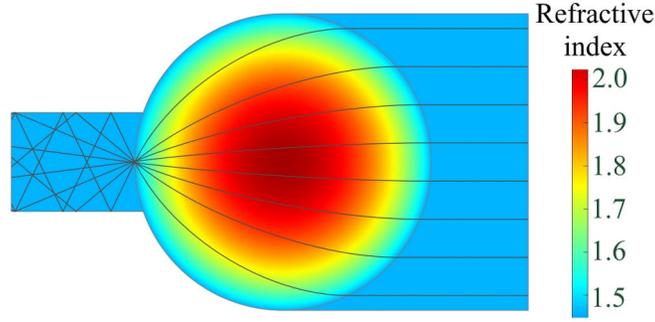

Fig. 1. Coupling a wider waveguide to a narrower waveguide by Luneburg lens.

## 3. Multilayered Luneburg lens

Different methods have been used to implement GRIN lenses such as graded photonic crystal (GPC) [25, 26], varying the thickness of the guiding layer in a slab waveguide [27, 28], and multilayer structures [19]. In this paper, we implement the Luneburg lens as a ring-based multilayer structure. Different applications have been introduced based on the interference effect in multilayer structures [29-32]. However, the interference effect diminishes considerably as the thickness of the layers become much smaller than the wavelength of the light. A multilayer structure with subwavelength layer thicknesses is regarded as an anisotropic homogeneous medium [33]. For transverse magnetic (TM) mode where the electric field is parallel to the inclusion layers, the refractive index of the multilayer structure is approximated by [34]:

$$n^2_{eff,TM} = f_{inc} n^2_{inc} + (1 - f_{inc}) n^2_{host} \quad (2)$$

where $f_{inc}$ is the fraction of the total volume occupied by inclusion layer. The $n_{host}$, $n_{inc}$, and $n_{eff,TM}$ are the refractive indices of the host, inclusion, and effective medium for TM mode, respectively. The host material is considered to be the same as the silica waveguide's core while the inclusion layers are silicon. The Luneburg lens with $R_{lens}$=1395 nm is divided into 9 annular rings with equal widths of $\Lambda$=155 nm. The width of inclusion layer in the i-th layer, $w_i$, is calculated by [19]

$$w_i = \frac{n^2_{eff,TM} - n^2_{host}}{n^2_{inc} - n^2_{host}} \Lambda \quad (3)$$

where $n_{eff,TM}$ is the average refractive index of the layer. The implemented ring-based multilayer structure is shown in Fig. 2(a). In this implementation the width of inclusion layer near to the edge of the lens is about 12 nm which is difficult to manufacture. Inspired by GPC, we limit the width of inclusion layers to $w_{min}$=35 nm which is shown in Figs. 2 (b) and (c). Limiting the minimum width of inclusion layers is achieved by dividing each annular ring into smaller annular sectors. Then for each annular sector, the arc length of the inclusion sector with a given width, $w_{min}$, is calculated. The length of the inclusions is also larger than 35 nm. The lens implemented by this method is displayed in Fig. 2(b). As shown in Fig. 2(c), the lens is truncated by a parabolic function to improve the performance of the lens which is discussed in next section.

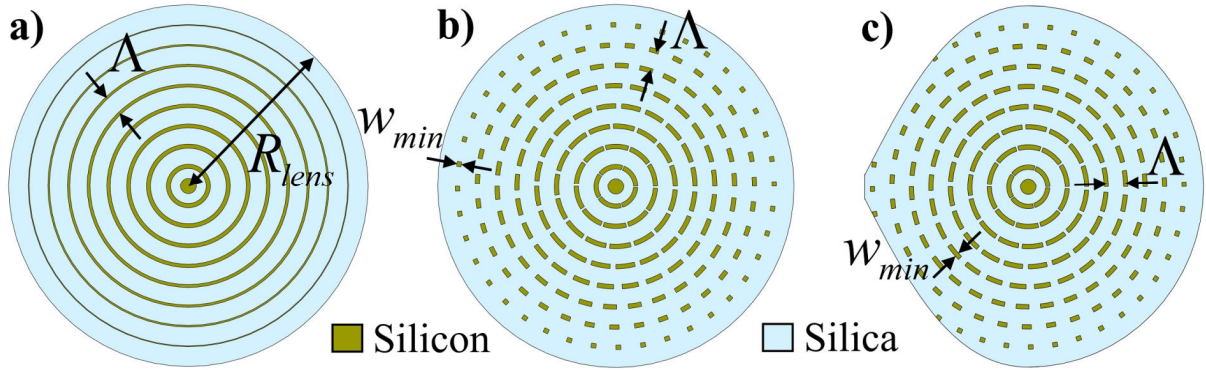

Fig. 2. Multilayered Luneburg lens with $R_{lens}$=1395 nm and $\Lambda$=155 nm. a) Simple concentric cylindrical multilayer structure. The width of silicon layer near to the edge of the lens is about 12 nm. b) The width of the inclusion layers (i.e., silicon layers) are limited to $w_{min}$=35 nm. c) The lens is truncated with a parabolic function.

## 4. Results and discussion

The PhC structure considered here is a two-dimensional (2D) square array of silicon rods with a lattice constant of $a$=465 nm surrounded by silica. The radius of rods is $r$=0.2$a$=93 nm [35]. This PhC structure has a bandgap in TM mode where the electric field is parallel with the axis of the rods. This bandgap covers wavelengths in free space from 1440 to 1680 nm [35]. Similar to previous studies [11-15, 36], we utilize 2D simulations to evaluate the performance of the designed coupler. Finite element method (FEM) is used to calculate the scattering parameters. Simulations are performed with Comsol Multiphysics®. We also consider that the silica waveguide is surrounded by air [12, 15, 37]. The width of the silica waveguide is $W_{WG}$=6$a$=2.79 μm while the length of the coupler is $L$=2$R_{lens}$-$r$=2.7 μm. The structure of the coupler and designing parameters are shown in Fig. 3. The coupler tip width is denoted by $W_{tip}$. Ideally, the center of the coupler's tip should match with the center of PhC waveguide, however, alignment error may occur which is shown by $d_{misalign}$. The "Free Triangular" mesh is used with a maximum element size of 100 nm while the minimum element size is determined by Comsol. When there are a large number of small objects, in this case rods, the meshing takes a long time. In Comsol, the meshing time can be reduced considerably by coping the mesh in the periodic structures. Hence, we mesh a single unit cell of the PhC and then copy it to the rest of the PhC structure. Another technique to reduce the meshing time of the complicated structures is to divide the simulation domain into smaller domains and then mesh them separately. Here, we first mesh the inclusion layers and then the remaining domains are meshed. The ports are used to evaluate the scattering parameters are also shown in Fig. 3. Terminating the PhC waveguide with conventional perfectly matched layer (PML) results in considerable spurious reflection [38, 39]. This spurious reflection introduces large errors in scattering parameter calculations. To overcome this problem, the PhC waveguide is terminated by distributed-Bragg-reflector [39], or PhC-based [38] PML domain. In the PhC-based PML, each waveguide is truncated with a port and a homogeneous domain of a PML surrounded by a few periods of the PhC lattice as shown in Fig. 3. The width of PML domains is 4×$a$. In Comsol, the slit condition should be applied to the interior ports. The electromagnetic field is almost restricted to the waveguides, therefore, scattering boundary condition (SBC) is applied for the remaining computational boundaries without introducing any spurious reflection.

The electric field distribution of the TM mode light through the coupler is shown in Fig. 4. In this figure, the designing parameters are $W_{tip}= 2a-8r_{rod}=186$ $nm$, $d_{misalign}=0$, and $d_{cut}=1.0\times r_{rod}=93$ $nm$. The return and coupling losses at a wavelength of 1550 nm are 19.2 and 0.40 dB, respectively. The performance of the complete and truncated couplers are compared in subsection 4.1. The dependence of coupling efficiency on the cut position of the PhC structure, $d_{cut}$, is studied in subsection 4.2. The effect of the coupler tip width, $W_{tip}$, on the coupling efficiency is discussed in subsection 4.3. The coupling loss introduced by alignment error, $d_{misalign}$, of the coupler is examined in subsection 4.4. We also compare our results with previous studies in subsection 4.5.

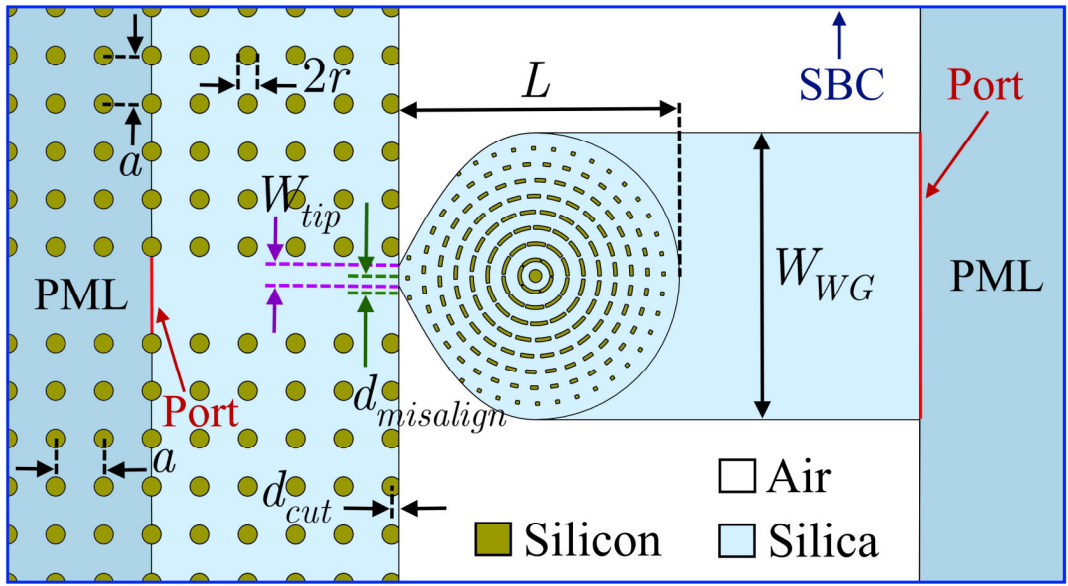

Fig. 3. The truncated Luneburg lens as a coupler and designing parameters.

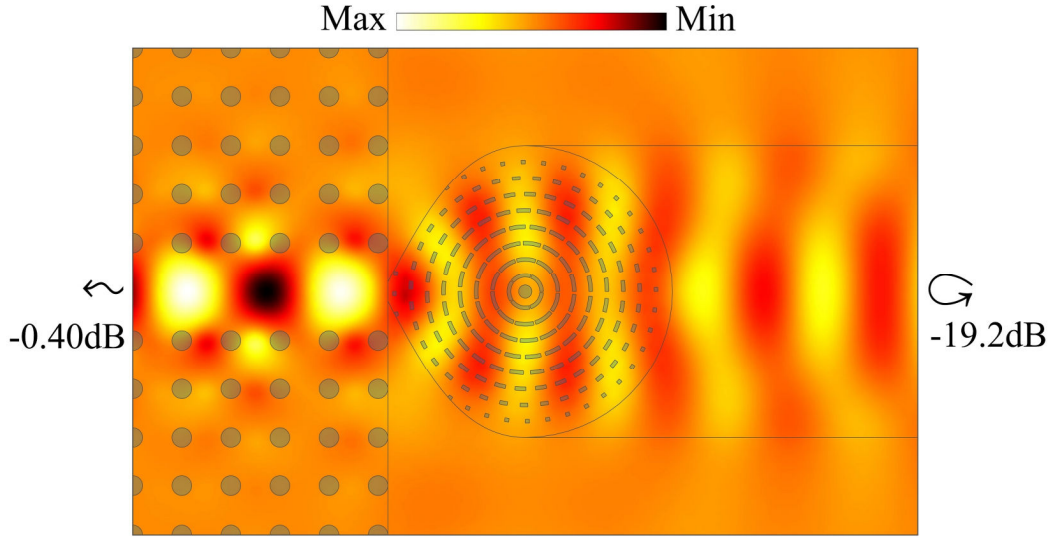

Fig. 4. The electric field distribution of the TM mode light at 1550 nm through the coupler with designing parameters of $W_{tip}$= 2a-8$r_{rod}$=186 nm, $d_{misalign}$=0 nm, $d_{cut}$=1.0×$r_{rod}$=93 nm.

## 4.1 Comparison of complete vs truncated Luneburg coupler

The performance of the Luneburg lens as a coupler in its complete and truncated forms is compared in Fig. 5. The coupling efficiencies of the designed couplers are evaluated for connecting a silica waveguide to a PhC waveguide and vice versa. The coupling loss of the complete lens is lower than 1.67 dB. However, the truncated lens benefits from the tapering effect, therefore, its performance is improved compared to the complete lens. For the truncated Luneburg coupler, the coupling loss is lower than 0.49 dB in the C-band while in the S, C, L, and U-bands the coupling loss is lower than 1.28 dB.

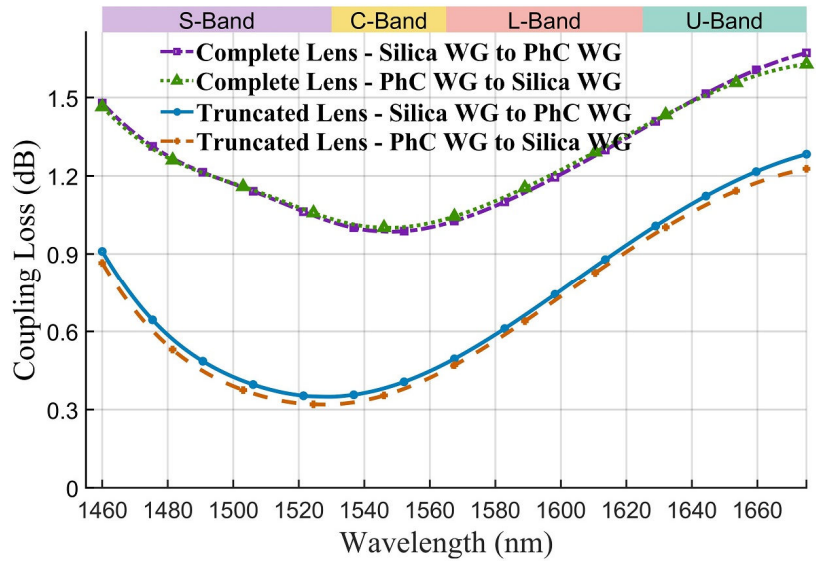

Fig. 5. Efficiency of the complete and truncated Luneburg lenses in coupling a silica waveguide (WG) into a PhC waveguide and vice versa.

## 4.2 Cut position of PhC

The coupling efficiency is strongly affected by the cut position of the PhC structure due to the modal properties of the Bloch modes in the PhC waveguide [40]. The average coupling loss in the C-band based on the cut position of the PhC structure is shown in Fig. 6. As the cut position ($d_{cut}$) decreases, larger portion of the optical wave is transformed into a surface wave, i.e., the electromagnetic wave propagates at the interface of the PhC structure and a homogenous medium (air) [41]. For the cut position of $d_{cut}=0$, the surface wave reaches its maximum magnitude. On the other hand, as $d_{cut}$ increases, the magnitude of the surface wave decreases resulting in the reduction of coupling loss. The average coupling loss in the C-band reaches its minimum at $d_{cut}=1.0 \times r_{rod}$. In these simulations, $d_{misalign}=0$ and $W_{tip}= 2a-8r_{rod}=465$ $nm$ are considered.

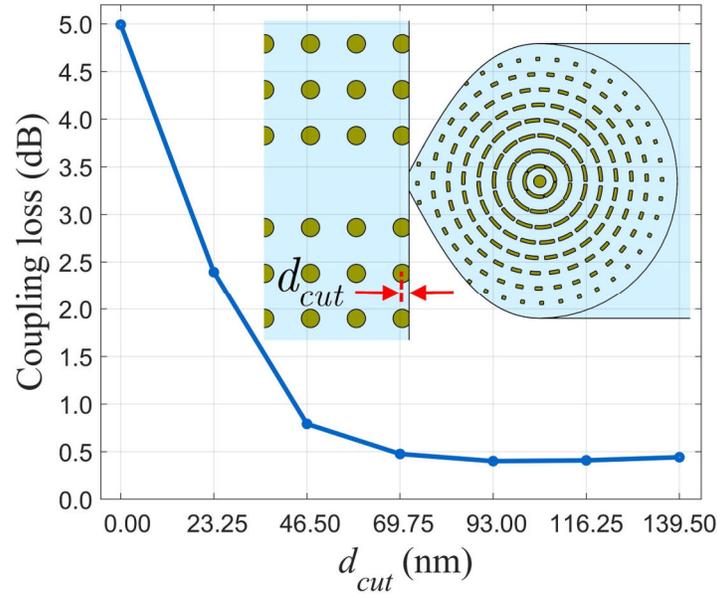

Fig. 6. Average coupling loss in the C-band vs cut position of PhC

## 4.3 Coupler tip width

The coupler tip width, $W_{tip}$, affects the coupling efficiency of the designed coupler. When there is no misalignment error, $d_{misalign}=0$, and the cut position is $d_{cut}=1.0 \times r_{rod}=93$ nm, we optimize the coupler tip width to minimize the coupling loss. The average coupling loss in the C-band based on different coupler tip width is shown in Fig. 7. As $W_{tip}$ approaches the width of the PhC waveguide, $2a$, the slope of the parabolic function used to truncate the lens decreases and consequently its tapering effect decreases. Therefore, the coupling loss increases as $W_{tip}$ increases. The minimum coupling loss in the C-band is achieved for $W_{tip}= 2a-8r_{rod}=186$ $nm$.

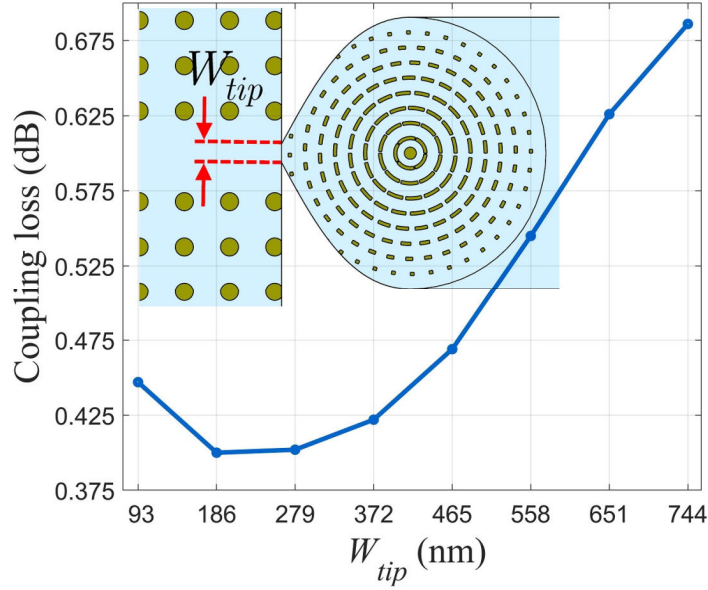

Fig. 7. Average coupling loss in the C-band vs coupler tip width

### 4.4 Misalignment and fabrication tolerance

In order to minimize the coupling loss, the tip of the coupler should be aligned with the center of the PhC and silica waveguides. We consider the alignment error of the coupler with respect to the center of the PhC waveguide denoted by $d_{misalign}$ in Fig. 3. The average coupling loss in the C-band due to the alignment error is displayed in Fig. 8. Obviously, as the alignment error increases the coupling loss increases. However, the degradation of the coupler's performance is relatively low with respect to alignment error. The average coupling loss in the C-band remains below 2.3 dB for alignment errors up to $d_{misalign}=2.0 \times r_{rod}=186$ nm.

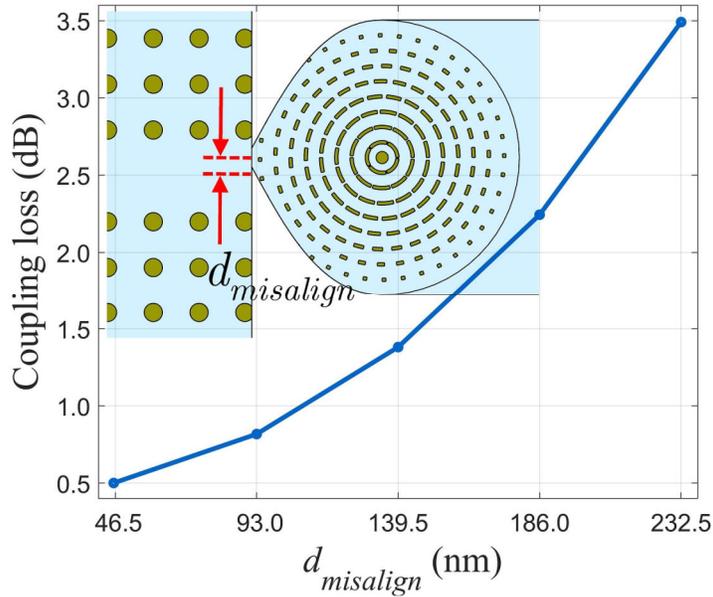

Fig. 8. Average coupling loss in the C-band vs misalignment

Some deviations are expected in the fabrication process of the designed coupler. We numerically estimate the effect of fabrication imperfections on the performance of the designed coupler. To this end, we consider three deviations in the fabrication of silicon inclusions: variation in width and length, displacement, and smoothed edges. We introduce a random deviation of ±20 nm in the width/length of the inclusions. And random displacements of ±20 nm in both the x and y-directions are introduced to the position of the inclusions. Finally, the sharp edges of the inclusions are randomly smoothed with fillet radius of 5-20 nm. The numerical simulations indicate that introducing these imperfections result in a maximum excess loss of 0.2 dB in the C-band.

### 4.5 Comparison with previous studies

We compare our design with previous studies in Table 4. The coupling mechanism, PhC lattice structure, width of the dielectric waveguide, and coupler's length are compared in this table. The transmission efficiency of the truncated Luneburg lens is also compared with previous studies in Fig. 9. Reference [11] is designed for the range of 900 nm, so it is not presented in this figure. References [12, 14] have high transmission efficiencies, however, they have narrower bandwidth compared to our design. Here, we compare the 9.69 µm-long convex taper of reference [16] since it is in the same range as the dielectric waveguide's width. However, 18.24 µm-long convex, concave, and linear nonuniform PhC tapers have also been presented in [16] with higher transmission efficiencies. In this method, the lattice structure is modified, therefore, it could not be used for lattices with large scatterers (holes or rods). For large scatterers, the scatterers may overlap leading to the destruction of the photonic bandgap [17]. We also compare our results with silicon wire-to-PhC waveguide couplers [42, 43]. In reference [42], a PhC taper is optimized to match the grouping index and mode size. The coupling loss of this design is lower than 0.5 dB for the wavelength range of 12 nm in the C-band. In reference [43], a mode converter is utilized to couple a strip waveguide to a slot PhC waveguide. The measured coupling loss of 0.08 dB is achieved in the bandwidth of 1520-1580 nm, however, the length of the coupler is 30 µm. Our designed coupler has a wider bandwidth with a transmission efficiency of higher than 75%, therefore, it has an increased potential for various applications and facilitates signal detection at the output [10].

Table 4. Comparison of transmission efficiency of truncated Luneburg coupler with previous studies

| Ref. | Coupling Mechanism | PhC Lattice Type | Width of dielectric waveguide (µm) | Length of coupler (µm) |
|---|---|---|---|---|
| [11] | PhC step taper | Hole-type triangular | 1.6 | 3 |
| [12] | PhC taper -Longitudinal gliding of lattices | Rod-type square heterostructure | 2.5 | 2.4 |
| [13] | PhC taper with a defect | Hole-type triangular | 3 | 2.3 |
| [14] | PhC taper with defects | Rod-type triangular | 3 | 2.79 |
| [15] | PhC taper with defects | Rod-type triangular | 3 | 1 |
| [16] | Nonuniform PhC taper | Rod-type square | 10 | 9.69 |
| [42] | Optimized PhC taper | Hole-type triangular | 0.86 | 2.4 |
| [43] | Mode converter | Hole-type triangular | 0.45 | 30 |
| Luneburg coupler | Luneburg lens | Rod-type square | 2.79 | 2.7 |

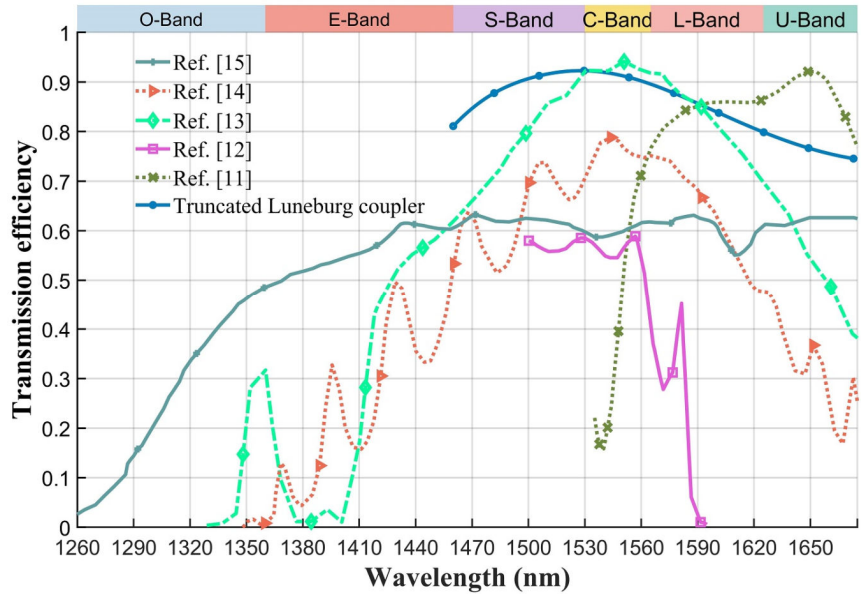

Fig. 9. Comparison of the designed coupler's transmission efficiency with previous studies.

## 5. Conclusion

We present and numerically study a silica waveguide to PhC waveguide coupler based on the focusing property of the Luneburg lens. Truncating the Luneburg lens in the shape of a parabolic taper improves its coupling efficiency. The average coupling loss between a 2.79 μm-wide silica waveguide and a single-line defect PhC structure with a rod-type square lattice is 0.40 dB in the C-band. In the entire S, C, L, and U bands of optical communications, the coupling loss is lower than 1.28 dB. The length of the designed coupler is 2.7 μm which is implemented by a concentric cylindrical multilayer structure. In order to avoid inclusion layers with very narrow widths in the multilayer structure, we divide each layer into a number of layers with defined minimum width.